# Quantum theory of a Bose-Einstein condensate out of equilibrium

Aranya B Bhattacherjee[1], Vikash Ranjan[2] and ManMohan[2]

[1]Department of Physics, A.R.S.D College (University of Delhi, South Campus), Dhaula Kuan, New Delhi-110 021, India
[2]Department of Physics and Astrophysics, University of Delhi, Delhi-110 007, India.

**Abstract**: We consider the interaction between a single-mode quantized perturbing external field and a Bose-Einstein condensate (BEC) out of equilibrium. The usual Rabi type oscillations between the ground and the excited state of the coherent topological modes are observed with a Rabi frequency modified by the two-body atomic interactions. Taking into account the granular structure of the external perturbing field reveals the well know phenomena of collapse and revival of the Rabi oscillations. In particular we find that atomic interactions reduce the Rabi frequency and also affect the collapse and revival sequence.

## Introduction

Bose-Einstein condensation of trapped atoms can be understood as the condensation to the ground coherent state, that is, to the state with the lowest single-particle energy. In an equilibrium statistical system, the Bose-condensate state is always the ground single particle state.

An intriguing question is whether one could create non-ground state condensates of Bose atoms, that is a macroscopic occupation of a non-ground single particle state. Clearly, this could only be done in a non-equilibrium system. If an additional field is switched on, with a frequency in resonance with the transition frequency between two coherent energy levels, then higher topological modes can be excited, thus creating non-ground state condensates. Coherent states of trapped atoms are the solutions of the non-linear Schrödinger equation with a confining potential. Stationary solutions of this equation form a discrete set of wavefunctions, corresponding to different energy levels, are called coherent modes. Such topological modes were first proposed by Yukalov et.al. [2], these condensates are associated with excited coherent modes. For instance, a rotating field for exciting vortices can be created by multiple laser fields [3-5]. The resonant excitation can lead to the formation of several vortices [6]. Topological coherent modes have also been studied in [7,8]. Anomalous dynamical behavior of these coherent modes was found in [9,10]. An explanation for this anomalous behavior was put forward in [11]. The aim of the present paper is to develop a preliminary quantum theory for the interaction of quantized electromagnetic field and two-level ultra-cold Bosonic atoms out of equilibrium. An early work of Zhang [12] based on a similar model reported inter-atom interactions included by light photon exchanges between ultra-cold atoms. We study population inversion dynamics in this non-equilibrium system, using a formalism similar to that used in quantum optics.

## The Model

Consider a gas of Bosonic two-level atoms, which interact with a weak, quantized external perturbing field. In an equilibrium statistical system, the Bose-condensate is always the ground single-particle state. In a non-equilibrium system, it should be possible to create excited condensates of Bose-atoms, i.e. a macroscopic occupation of a non-ground single-particle state. In order to transfer atoms from a single particle ground state, with energy $E_0$, to another of higher state $E_e$, one needs a near resonant transition, i.e. the frequency of the external perturbing field should be close to $(E_e - E_0)/\hbar$. Our theory begins with the second-quantized Hamiltonian

$$H = H_{atom} + H_{field} + H_{atom-field} + H_{atom-atom} \tag{1}$$

Where $H_{atom}$ and $H_{field}$ give the free evolution of the atomic and perturbing fields respectively. $H_{atom-field}$ describes the coupling between the atomic field and external perturbing field. $H_{atom-atom}$ contains the two-body s-wave scattering collisions between atoms. The free atomic Hamiltonian is given by

$$H_{atom} = \int d^3\vec{r} \begin{bmatrix} \psi_g^+(\vec{r})\left(-\frac{\hbar^2\vec{\nabla}^2}{2m} + \hbar\omega_g + V_g(\vec{r})\right)\psi_g(\vec{r}) + \\ \psi_e^+(\vec{r})\left(-\frac{\hbar^2\vec{\nabla}^2}{2m} + \hbar\omega_e + V_e(\vec{r})\right)\psi_e(\vec{r}) \end{bmatrix} \quad (2)$$

Where $m$ is the atomic mass. $\hbar\omega_g$ and $\hbar\omega_e$ are the atomic energies of the ground and the excited level. $V_g(\vec{r})$ and $V_e(\vec{r})$ are the trap potentials of the ground and the excited states. The free evolution of the external field is governed by the Hamiltonian

$$H_{field} = \hbar\omega_a a^+ a \quad (3)$$

$a$ and $a^+$ are the annihilation and creation operators of the external field. The atomic and external perturbing fields interact in the dipole approximation via the Hamiltonian

$$H_{atom-field} = \hbar g \int a\psi_g(\vec{r})\psi_e^+(\vec{r})d^3\vec{r} + \hbar g \int a^+\psi_g^+(\vec{r})\psi_e(\vec{r})d^3\vec{r} \quad (4)$$

Where $g$ is the atom-field coupling constant. Finally the collision Hamiltonian is taken to be

$$H_{atom-atom} = \frac{2\pi\hbar^2 a_{gg}}{m}\int d^3\vec{r}\psi_g^+(\vec{r})\psi_g^+(\vec{r})\psi_g(\vec{r})\psi_g(\vec{r})$$

$$\frac{2\pi\hbar^2 a_{ee}}{m}\int d^3\vec{r}\psi_e^+(\vec{r})\psi_e^+(\vec{r})\psi_e(\vec{r})\psi_e(\vec{r}) + \frac{4\pi\hbar^2 a_{ge}}{m}\int d^3\vec{r}\psi_g^+(\vec{r})\psi_g(\vec{r})\psi_e^+(\vec{r})\psi_e(\vec{r})$$

(5)

Here $a_{ij}$ is the s-wave scattering length in the states $i,j$. Here we have assumed that $a_{eg} = a_{ge}$. The study of the quantum statistical properties of the condensate (at T=0) can be reduced to a relatively simple model by using a mode expansion and subsequent truncation to just a single mode (the "condensate mode"). In particular, one writes the Heisenberg ground state atomic field annihilation operator as a mode expansion over single particle states,

$$\psi_g(\vec{r},t) = \sum_\alpha b_\alpha(t)\phi_\alpha^g(\vec{r})\exp(-i\mu_\alpha t/\hbar) = b_0(t)\phi_0^g(\vec{r})\exp(-i\mu_0 t/\hbar) + \tilde{\psi}(\vec{r},t)$$

(6)

Where $\{\psi_\alpha(\vec{r})\}$ are a complete orthonormal basis set and $\{\mu_\alpha\}$ the corresponding eigenvalues. The first term in the second line of eqn.(6) acts only on the condensate state vector, with $\phi_0^g(\vec{r})$ chosen as a solution of the stationary Gross-Pitaeviskii equation. The second term, $\tilde{\psi}(\vec{r},t)$, accounts for non-condensate atoms. In a similar manner, we can expand the excited state atomic field annihilation operator

$$\psi_e(\vec{r},t) = \sum_\beta c_\beta(t)\phi_\beta^\alpha(\vec{r})\exp(-iv_\beta t/\hbar) \qquad (7)$$

Retaining only the lowest mode of the ground state and the excited state, we can rewrite eqns.(2)-(5).

$$H = \hbar\omega_a a^+ a + \hbar\omega_b b_0^+ b_0 + \hbar\omega_c c_0^+ c_0 + \frac{\hbar\kappa_{gg}}{2}b_0^+ b_0^+ b_0 b_0$$
$$+ \frac{\hbar\kappa_{ee}}{2}c_0^+ c_0^+ c_0 c_0 + \hbar\kappa_{eg}b_o^+ b_0 c_0^+ c_0 + \hbar g G\left[ab_0 c_0^+ + a^+ c_0 b_0^+\right] \qquad (8)$$

Where,

$$\hbar\omega_b = \int d^3\vec{r}\left[\phi_0^{*g}(\vec{r})\left(-\frac{\hbar^2\vec{\nabla}^2}{2m} + \hbar\omega_g + V_g(\vec{r})\right)\phi_0^g(\vec{r})\right]$$

$$\hbar\omega_e = \int d^3\vec{r}\left[\phi_0^{*e}(\vec{r})\left(-\frac{\hbar^2\vec{\nabla}^2}{2m} + \hbar\omega_e + V_e(\vec{r})\right)\phi_0^e(\vec{r})\right]$$

$$\hbar\kappa_{gg} = \frac{4\pi\hbar^2 a_{gg}}{m}\int d^3\vec{r}\left|\phi_0^g(\vec{r})\right|^4$$

$$\hbar\kappa_{ee} = \frac{4\pi\hbar^2 a_{ee}}{m}\int d^3\vec{r}\left|\phi_0^e(\vec{r})\right|^4$$

$$\hbar\kappa_{eg} = \frac{4\pi\hbar^2 a_{eg}}{m}\int d^3\vec{r}\left|\phi_0^e(\vec{r})\right|^2\left|\phi_0^g(\vec{r})\right|^2$$

$$\hbar G = \hbar g\int \phi_0^g(\vec{r})\phi_0^{*e}(\vec{r})d^3\vec{r} = \hbar g\int \phi_0^e(\vec{r})\phi_0^{*g}(\vec{r})d^3\vec{r}$$

Here $g$ is assumed to be real. The Hamiltonian of eqn.(8) can be rewritten as

$$H = \hbar\omega_a a^+ a + \frac{1}{2}\hbar\omega_0 \hat{\sigma}_z + \hbar\gamma_0 \hat{\sigma}_z + \hbar\gamma_1 \hat{\sigma}_z^2 + \hbar G(\hat{\sigma}_+ a + a^+ \hat{\sigma}_-) \quad (9)$$

Where, $\hbar\omega_0 = \hbar(\omega_b - \omega_c)$, $\hat{\sigma}_{b_0 b_0} + \hat{\sigma}_{c_0 c_0} = 1$, $\hat{\sigma}_{b_0 b_0} = b_0^+ b_0$, $\hat{\sigma}_{c_0 c_0} = c_0^+ c_0$, $\hat{\sigma}_z = \hat{\sigma}_{b_0 b_0} - \hat{\sigma}_{c_0 c_0}$, $\hat{\sigma}_+ = b_0 c_0^+$, $\hat{\sigma}_- = c_0 b_0^+$, $\gamma_0 = (\kappa_{gg} - \kappa_{ee})/4$, $\gamma_1 = (2\kappa_{gg} + 2\kappa_{ee} - \kappa_{eg})/16$.

## Population dynamics

The Hamiltonian of eqn.(9) is the variant of the Jaynes-Cummings model in quantum optics. $\hat{\sigma}_\pm$ and $\hat{\sigma}_z$ satisfy the spin-1/2 algebra of the Pauli matrices,

$$[\hat{\sigma}_-, \hat{\sigma}_+] = -\hat{\sigma}_z, \quad [\hat{\sigma}_-, \hat{\sigma}_z] = 2\hat{\sigma}_- \quad (10)$$

The Heisenberg equations for the operators $a$, $\hat{\sigma}_-$ and $\hat{\sigma}_z$ are obtained from the Hamiltonian (9),

$$\dot{a} = \frac{1}{i\hbar}[a, H] = -i\omega_a a - iG\hat{\sigma}_- \quad (11)$$

$$\dot{\hat{\sigma}}_- = -i\tilde{\omega}\hat{\sigma}_- + iGa\hat{\sigma}_z \quad (12)$$

$$\dot{\hat{\sigma}}_z = 2iG(a^+\hat{\sigma}_- - a\hat{\sigma}_+) \quad (13)$$

Where $\tilde{\omega} = \omega_0 + 2\gamma_0$. In order to facilitate a solution of these coupled operator equations, we define the following constants of motion,

$$N = a^+ a + \hat{\sigma}_+ \hat{\sigma}_- \quad (14)$$

$$C = \frac{\Delta' \hat{\sigma}_z}{2} + G(\hat{\sigma}_+ a + a^+ \hat{\sigma}_-) \quad (15)$$

$\Delta' = \Delta + 2\gamma_0$ and the detuning $\Delta = \omega_0 - \omega_a$. $N$ and $C$ commute with $H$. The desired equations of motion for the atomic lowering operator $\hat{\sigma}_-$ and photon annihilation operator $a$ are

$$\ddot{\hat{\sigma}}_- + 2i(\omega_a - C)\dot{\hat{\sigma}}_- + (2\omega_a C - \omega_a^2 + G^2)\hat{\sigma}_- = 0 \quad (16)$$

$$\ddot{a} + 2i(\omega_a - C)\dot{a} + (2\omega_a C - \omega_a^2 + G^2)a = 0 \quad (17)$$

These equations can be solved in a straightforward manner and the resulting expressions for $\hat{\sigma}_-(t)$ and $a(t)$ are

$$\hat{\sigma}_-(t) = [\hat{\sigma}_+(t)]^\dagger = \exp(-i\omega_a t)\exp(iCt)\left[\begin{array}{l}\left(\cos(Rt) + iC\dfrac{\sin(Rt)}{R}\right)\hat{\sigma}_-(0) \\ -iG\dfrac{\sin(Rt)}{R}a(0)\end{array}\right] \quad (18)$$

$$a(t) = \exp(-i\omega_a t)\exp(iCt)\left[\begin{array}{l}\left(\cos(Rt) - iC\dfrac{\sin(Rt)}{R}\right)a(0) \\ -iG\dfrac{\sin(Rt)}{R}\hat{\sigma}_-(0)\end{array}\right] \quad (19)$$

Here the Rabi frequency $R = \left[\dfrac{\Delta^2}{4} + G^2\right]^{1/2}$. The expressions for the ground state population ($\hat{\sigma}_{b_0 b_0}$), excited state populations ($\hat{\sigma}_{c_0 c_0}$) and population inversion ($W(t)$) are found to be

$$\hat{\sigma}_{b_0 b_0} = 1 - \dfrac{G^2 \sin^2(Rt/2)}{R^2} = 1 - \dfrac{\sin^2\left(\left[1 + \left(\dfrac{\Delta}{2G} + \dfrac{\gamma_0}{G}\right)^2\right]^{1/2} Gt/2\right)}{\left[1 + \left(\dfrac{\Delta}{2G} + \dfrac{\gamma_0}{G}\right)^2\right]} \quad (20)$$

$$\hat{\sigma}_{c_0 c_0} = \dfrac{G^2}{R^2}\sin^2(Rt/2) = \dfrac{\sin^2\left(\left[1 + \left(\dfrac{\Delta}{2G} + \dfrac{\gamma_0}{G}\right)^2\right]^{1/2} Gt/2\right)}{\left[1 + \left(\dfrac{\Delta}{2G} + \dfrac{\gamma_0}{G}\right)^2\right]} \quad (21)$$

$$W(t) = <\hat{\sigma}_z> = [|b_0|^2 - |c_0|^2] = 1 - \dfrac{2G^2 \sin^2(Rt/2)}{R^2} = 1 - \dfrac{2\sin^2\left(\left[1 + \left(\dfrac{\Delta}{2G} + \dfrac{\gamma_0}{G}\right)^2\right]^{1/2} Gt/2\right)}{\left[1 + \left(\dfrac{\Delta}{2G} + \dfrac{\gamma_0}{G}\right)^2\right]} \quad (22)$$

These expressions are similar to those found in Ref. [2] and reviewed in Ref. [1]. To understand the effect of atomic interactions on the dynamics of population inversion operator $W(t)$, we have plotted in fig.1 the time evolution (in the units of Rabi frequency) of $W(t)$ for three different values of $\alpha = \gamma_0/G$. Clearly, we observe that when $\alpha = 0$, i.e in the absence of atomic interactions, complete population transfer between the excited and ground state takes places. With increasing $\alpha$, the Rabi oscillations seem to die out and the entire atomic population is trapped in the initial state. Actually, the origin of this behavior is the shifting of the original unperturbed ground and excited state energy levels away from each other due to atomic interactions.

The expression for $W(t)$ in equation (22) is for an initial vacuum field. If the field has '$n$' quanta present initially then the expression for $W(t)$ would depend on '$n$'.

$$W(t) = \langle g, \alpha | \sigma_z(t) | g, \alpha \rangle \tag{23}$$

Here we have assumed that the atom is initially in the ground state and the field is initially in the coherent state $|\alpha\rangle$. The fully quantized expression for $W(t)$ is

$$W(t) = \sum_n \rho_{nn}(0) \left[ 1 - \frac{2(1+n)\sin^2\left(\left[1+n+\left(\frac{\Delta}{2G}+\frac{\gamma_0}{G}\right)^2\right]^{1/2} Gt/2\right)}{\left[1+n+\left(\frac{\Delta}{2G}+\frac{\gamma_0}{G}\right)^2\right]} \right] \tag{24}$$

Where the new Rabi frequency is $R_n = \left[\frac{\Delta^2}{4} + G^2(1+n)\right]^{1/2}$ and $\rho_{nn}(0) = \langle n \rangle e^{-\langle n \rangle}/n!$ is the probability that there are '$n$' quanta present in the field at time $t = 0$. In figure 2, $W(t)$ is plotted as a function of the normalized time '$Gt$' for an initial coherent state and two different values of the parameter $\alpha$. Clearly we observe collapse and revival of the Rabi oscillations. The collapse occurs because the Rabi oscillations associated with different excitations have different frequencies and therefore become uncorrelated. The revivals are purely a quantum feature and occur because the correlation is restored after some time. The collapse time $t_c$ and the revival time $t_r$ can be shown to be

$$t_c \cong \frac{1}{2G}\left(1+\left[\frac{\Delta}{2G\sqrt{\langle n \rangle}}+\frac{\gamma_0}{G\sqrt{\langle n \rangle}}\right]^2\right)^{1/2} \tag{25}$$

$$t_r \cong \frac{2\pi m\sqrt{\langle n \rangle}}{G}\left(1+\left[\frac{\Delta}{2G\sqrt{\langle n \rangle}}+\frac{\gamma_0}{G\sqrt{\langle n \rangle}}\right]^2\right)^{1/2} \qquad (26)$$

Where '$m$' is an integer. Equation (26) shows that revivals take place at regular intervals. Also both the collapse time $t_c$ and the revival time $t_r$ increase with increasing atomic interaction (this is also evident from figure 2).

## Conclusions

In conclusion, we have presented a quantum theory for the resonant excitation of non-equilibrium dynamics of trapped Bose-Einstein condensate interacting with a quantized external perturbing field. The principally important results of the paper are the demonstration of the inhibition of Rabi oscillations due to atomic interactions, the occurrence of collapse and revival of the Rabi oscillations due to quantum nature of the perturbing field. The collapse and the revival time are found to be directly proportional to the atomic interactions.

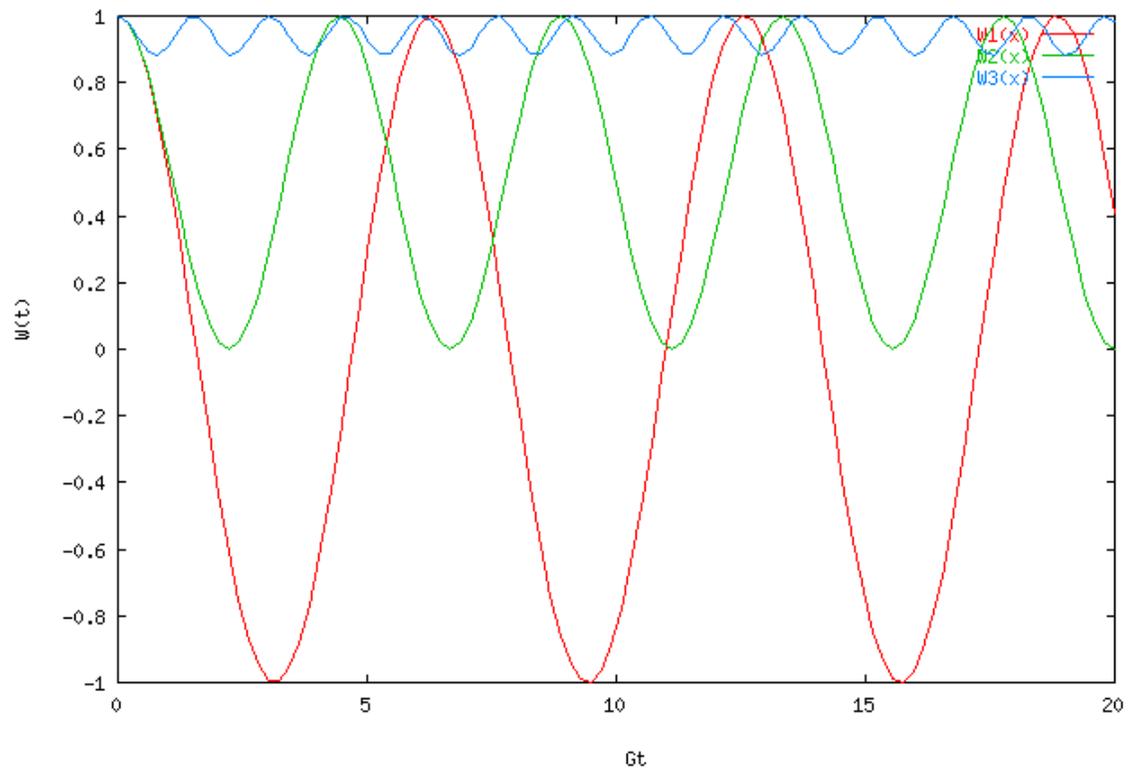

Figure 1: Time evolution (in units of $G$) of population inversion $W(t)$ for detuning $\Delta = 0$, $\alpha = 0$ (W1), $\alpha = 1$ (W2), $\alpha = 4$ (W3). Clearly, the effect of the atomic interactions is to inhibit the Rabi oscillations.

Figure 2

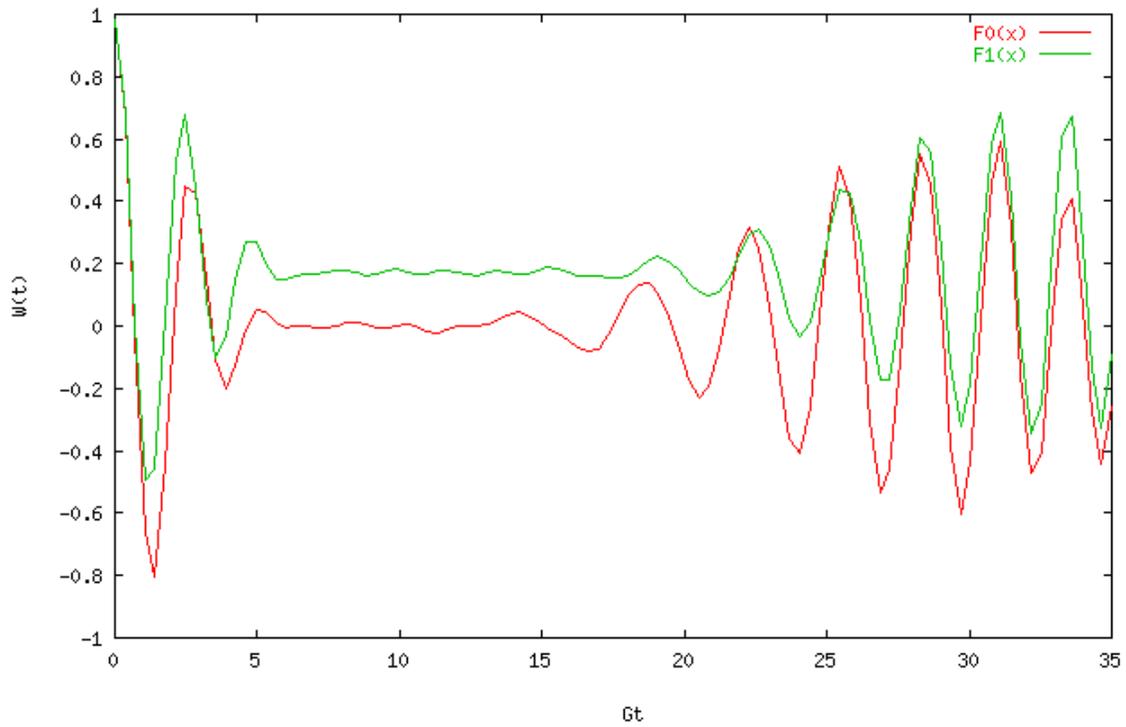

Figure 2: Time evolution (in units of $G$) of population inversion $W(t)$ for an initially coherent state for detuning $\Delta = 0$, $\alpha = 0$ (F0), $\alpha = 1$ (F1), $\langle n \rangle = 10$. Unlike Figure 1, here we observe the quantum phenomena of 'collapse and revival'.